# Considering the multi-time scale rolling optimization scheduling method of micro-energy network connected to electric vehicles

Hengyu Liu
*Electric Power Research·Institute of·State Grid Liaoning Electric Power Co., Ltd.*
*State Grid Liaoning Electric Power Supply·Co., Ltd.*
Shenyang, China
15840554655@163.com

Yanhong Luo
*School of Information Science and Engineering*
*Northeastern University*
Shenyang, China
luoyanhong@ise.neu.edu.cn

Congcong Wu
*School of New Energy and Electrical Engineering*
*Hubei University*
Hubei, China
ccwu@hubu.edu.cn

Yin Guan
*School of Energy and Power Engineering*
*Huazhong University of Science and Technology*
Wuhan, China
yinguan@hust.edu.cn

Ahmed Lotfy Elrefai
*Electrical Power Engineering*
*Egypt Japan University of Science and Technology*
Alexandria, Egypt
ahmed.lotfy@ejust.edu.eg

Andreas Elombo
*Namibia Energy Institute*
*Namibia University of Science and Technology*
Windhoek, Namibia
aelombo@nust.na

Si Li
*School of Information Science and Engineering*
*Northeastern University*
Shenyang, China
lisi@ise.neu.edu.cn

Sahban Wael Saeed Alnaser
*Department of Electrical Engineering*
*University of Jordan*
Amman, Jordan
s.alnaser@ju.edu.jo

Mingyu Yan
*School of Electrical and Electronic Engineering*
*Huazhong University of Science and Technology*
Wuhan, China
mingyuyan@hust.edu.cn

***Abstract*—The large-scale access of electric vehicles to the power grid not only provides flexible adjustment resources for the power system, but the temporal uncertainty and distribution complexity of their energy interaction pose significant challenges to the economy and robustness of the micro-energy network. In this paper, we propose a multi-time scale rolling optimization scheduling method for micro-energy networks considering the access of electric vehicles. In order to solve the problem of evaluating the dispatchable potential of electric vehicle clusters, a charging station aggregation model was constructed based on Minkowski summation theory, and the scattered electric vehicle resources were aggregated into virtual energy storage units to participate in system scheduling. Integrate price-based and incentive-based demand response mechanisms to synergistically tap the potential of source-load two-side regulation; On this basis, a two-stage optimal scheduling model of day-ahead and intra-day is constructed. The simulation results show that the proposed method reduces the scale of "preventive curtailment" due to more accurate scheduling, avoids the threat of power shortage to the safety of the power grid, and has more advantages in the efficiency of new energy consumption. At the same time, intra-day scheduling significantly reduces economic penalties and operating costs by avoiding output shortages, and improves the economy of the system in an uncertain forecasting environment.**

This work is supported by the State Grid Liaoning Electric Power Co., LTD. Science and Technology Project (No.2024YF-74), Research on key technologies of rural micro-energy network optimization control to support electric vehicles going to the countryside.

## I. INTRODUCTION

In response to global warming, many countries and regions have put forward carbon emission targets of "zero carbon" and "carbon neutrality" in recent years, and formulated a series of plans and policies to accelerate the global energy transition.

On the energy demand side, proactive and flexible demand-side management or demand-side response provides a solution [1]. Ref. [2] confirms that the demand side response should achieve a negative correlation between the renewable energy consumption rate and the total cost of the system. Ref. [3] proposes a CHPMEG coordinated optimization operation strategy considering load demand response and user thermal inertia to improve the efficiency of wind power consumption through an incentive mechanism.

Electric vehicles (EVs) have both elastic load and energy storage characteristics, and are high-quality demand response resources [4-5]. Ref. [6] eliminates the 12GW reserve capacity gap in the Portuguese grid through an EV charging load shifting algorithm; The microgrid management system constructed in Ref. [7] can simultaneously reduce costs and pollutant emissions. In order to solve the problem of insufficient capacity of a single EV [3], Ref. [8] proposes a time-to-space expansion model for charging stations to reduce the cost by 18% and the peak-to-valley difference by 31%. Ref. [9] established a zonotope-Minkowski quantitative system based on user behavior clustering to realize the evaluation of multi-dimensional scheduling potential.

For the optimal scheduling of integrated energy micro-energy grids, a multi-objective optimization model was developed based on the stepped carbon trading mechanism in Ref. [10] to improve the system economy. Considering the uncertainty of distributed energy resources, Ref. [11] proposes an operation strategy to reduce pollutant emissions and improve self-sufficiency. At the market level, the two-layer game model constructed in Ref. [12] significantly improves the renewable energy consumption rate. At the technical level, Ref. [13] enhances voltage stability through a network-load-storage co-optimization strategy. The above methods form a comprehensive optimization system covering the dimensions of economy, environmental protection and reliability.

In summary, the main contents of this article include:

(1) Based on Minkowski summation theory, the aggregation model of the charging station is constructed, the scattered electric vehicle resources are aggregated into virtual energy storage units, the dispatchable capacity of the charging station is calculated, and the maximum and minimum values of the dispatchable capacity are taken as the upper and lower limits of the virtual energy storage capacity constraints, respectively, to participate in the scheduling of the micro-energy network dispatching center.

(2) The demand response mechanism of the flexible load on the user side is introduced, and the incentive compensation measures are used to give full play to the flexible adjustment ability of the load side, which effectively suppresses the fluctuation of new energy output and realizes peak shaving and valley filling.

(3) A two-stage optimization scheduling strategy of day-ahead and intra-day optimization was established, which combined the day-ahead global optimization with intra-day rolling correction, and used CPLEX.The platform is used to solve the problem, which effectively improves the system in predicting the uncertain environment economy and robustness.

## II. MICRO-ENERGY NETWORK SYSTEM ARCHITECTURE

### A. *The overall architecture of the micro-energy network*

The overall architecture of the micro-energy network built in this paper is shown in the following figure:

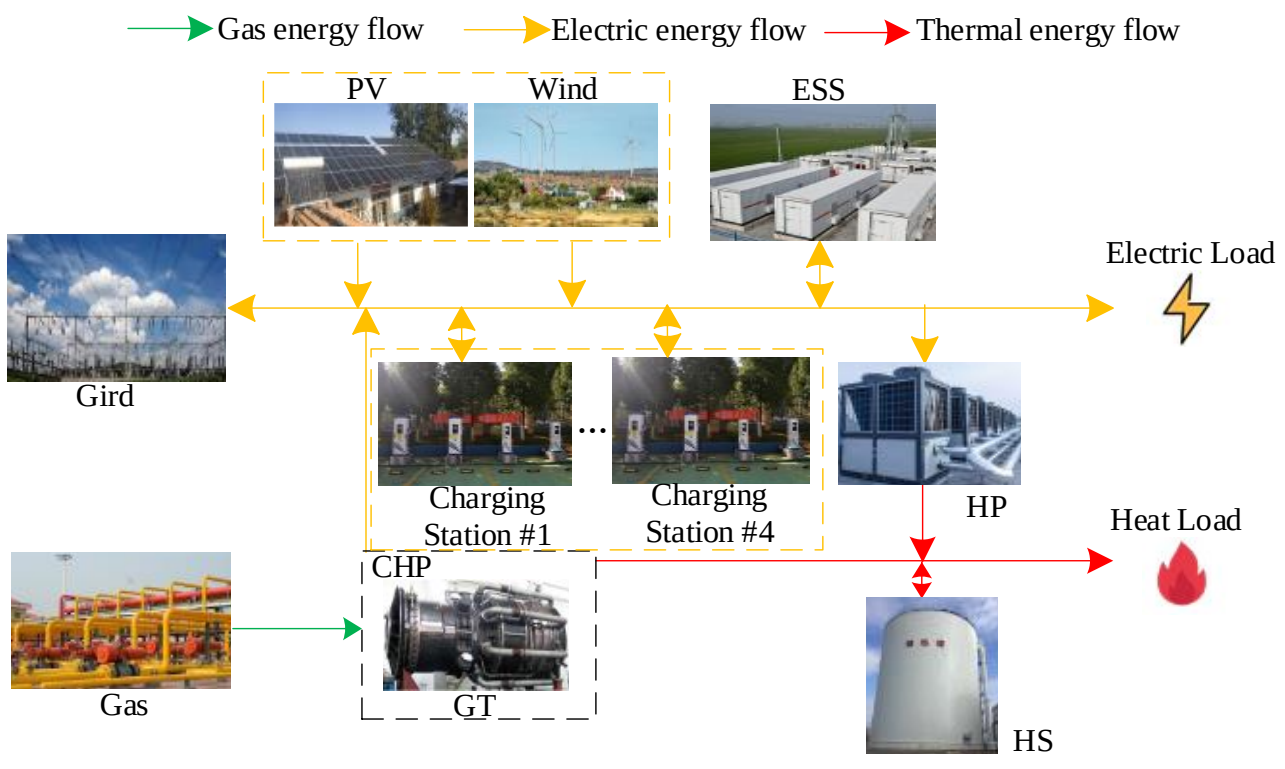

Fig. 1. Overall architecture of the microgrid.

### B. *Modeling of micro-energy grid system components*

(1) Liaison line transmission power model

$$P_{Gird_min} \le P_{Gird} \le P_{Gird_max} \tag{1}$$

(2) Micro gas turbine output model

$$P_{GT_min} \le P_{GE} \le P_{GT_max} \tag{2}$$

(3) Energy storage operation model

a) Power Model:

$$\begin{cases} 0 \le P_{ESS_ch}(t) \le P_{ESS_ch_max} \\ 0 \le P_{ESS_dis}(t) \le P_{ESS_dis_max} \\ P_{ESS_dis_max} = \min\left\{(SOC_{max} - SOC_{t-1}) \times C_{ESS} \times \eta_{dis}, P_{ESS_E}\right\} \\ P_{ESS_ch_max} = \min\left\{(SOC_{max} - SOC_{t-1}) \times C_{ESS} \times \dfrac{1}{\eta_{ch}}, P_{ESS_E}\right\} \end{cases} \tag{3}$$

where $C_{ESS}$ and $P_{ESS_E}$ represent the rated capacity and rated power of the battery, respectively.

b) Capacity Model:

$$\begin{cases} SOC_{min} \le SOC_t \le SOC_{max} \\ SOC_{start} = S_{end} \end{cases} \tag{4}$$

(4) Virtual energy storage operation model of charging station

a) Power Model:

$$\begin{cases} 0 \le P_{EVi_ch}(t) \le P_{EVi_ch_max} * P_{EVi_ch_s}(t) \\ 0 \le P_{EVi_dis}(t) \le P_{EVi_dis_max} * P_{EVi_dis_s}(t) \\ P_{EVi_dis_s}(t) + P_{EVi_ch_s}(t) \le 1 \end{cases} \tag{5}$$

where $P_{Evi_ch_s}(t)$ and $P_{Evi_dis_s}(t)$ represent the charging and discharging status of the charging station at its moment, respectively.

b) Capacity Model:

$$\begin{cases} SOC_{EVi_Forecast_min} \le SOC_{EVi_t} \le SOC_{EVi_Forecast_max} \\ SOC_{EVi_start} = SOC_{EVi_end} \end{cases} \tag{6}$$

$SOC_{Evi_t}$ is the capacity of the charging station $i$ at time $t$;

(5) Wind power photovoltaic output model

$$\begin{cases} 0 \le P_{pv}(t) \le N_{Pv} P_{pv0}(t) \\ 0 \le P_{wt}(t) \le N_{wt} P_{wt0}(t) \end{cases} \tag{7}$$

where $P_{wt0}(t)$ and $P_{Pv0}(t)$ represent the output power of a single wind turbine and photovoltaic panel;

## III. A TWO-STAGE OPTIMAL SCHEDULING MODEL OF MICRO-ENERGY NETWORK BASED ON THE CALCULATION OF DISPATCHABLE CAPACITY OF CHARGING STATIONS

### A. *Calculation of dispatchable capacity of electric vehicle charging station based on Minkowski summation theory*

(1) Dispatchable capacity of a single electric vehicle

The electric vehicle can participate in the scheduling of the micro-energy grid system as an energy storage only when it is in the grid-connected state, and in order to distinguish the off-grid state of the electric vehicle, the state variable $Dn,t$ is introduced, and it satisfies equation (8):

$$D_{n,t} = \begin{cases} 0, \forall t \notin \left[T_n^{arrive}, T_n^{leave}\right] \\ 1, \forall t \in \left[T_n^{arrive}, T_n^{leave}\right] \end{cases} \tag{8}$$

The control time area of the electric vehicle is extended to the whole optimal scheduling cycle, and the dispatchable capacity calculation model of a single electric vehicle is obtained: $0 \le P_{n,t}^{ch} \le P_{\max,n}^{ch} D_{n,t}, \forall n \in N_j^{EV}, \forall t \in T$ (9)

$$0 \le P_{n,t}^{dis} \le P_{\max,n}^{\text{dis}} D_{n,t}, \forall n \in N_j^{EV}, \forall t \in T \quad (10)$$

$$S_{n,t} = D_{n,t}\left(S_{n,t-1} + \eta^{ch} P_{n,t}^{ch} \Delta t - \frac{\eta^{ref} P_{n,t}^{dis} \Delta t}{\eta^{dis}}\right) \quad (11)$$

$$S_n^{\min} D_{n,t} \le S_{n,t} \le S_n^{\max} D_{n,t}, \forall n \in N_j^{EV}, \forall t \in T \quad (12)$$

where , $P_{n,t}^{ch}$、$P_{n,t}^{dis}$、$P_{\max,n}^{ch}$、$P_{\max,n}^{dis}$ , are the charging and discharging power of EVn and the maximum value of the power; $\eta_{ch}$, $\eta_{dis}$, and $\eta_{ref}$ represent the charging and discharging efficiency of the EV and the compensation coefficient for the discharge.

(2) The dispatchable capacity of the car charging station

The above equation (11) is converted so that the dispatchable model of a single electric vehicle has Minkowski additivity.

Eq. (11) is divided into three cases, firstly, the EV network access period, the initial power of EV needs to be considered $S_{n,\,arrive}$, corresponding to Eq. (13); followed by the normal grid-connected period of EV, corresponding to equation (14); Eq. (15) corresponds to the off-grid period of the EV, and the off-grid power of the EV needs to be considered $S_{n,leave}$.

$$S_{n,t} = S_{n,arrive} + \eta^{ch} P_{n,t}^{ch} \Delta t - \frac{\eta^{ref} P_{n,t}^{dis} \Delta t}{\eta^{dis}}, \quad \forall n \in N_j^{EV}, t = T_n^{arrive} \quad (13)$$

$$S_{n,t} = S_{n,t-1} + \eta^{ch} P_{n,t}^{ch} \Delta t - \frac{\eta^{ref} P_{n,t}^{dis} \Delta t}{\eta^{dis}}, \quad \forall n \in N_j^{EV}, \forall t = \left(T_n^{arrive}, T_n^{leave}\right] \quad (14)$$

$$S_{n,t} = S_{n,t-1} - S_{n,leave}, \quad \forall n \in N_j^{EV}, t = T_n^{leave} + \Delta t \quad (15)$$

And since $S_{n,t-1}=0$, $P_{n,t}^{ch} = 0$、$P_{n,t}^{\text{dis}} = 0$ Eq. (16) with Minkowski additivity:

$$\begin{cases} \begin{cases} t = T_n^{arrive} \\ t = T_n^{leave} + \Delta t \end{cases} \Rightarrow \begin{cases} D_{n,t}\left(D_{n,t} - D_{n,t-1}\right) = 1 \\ D_{n,t-1}\left(D_{n,t-1} - D_{n,t}\right) = 1 \end{cases} \\ S_{n,t} = S_{n,t-1} + S_{n,arrive} D_{n,t}\left(D_{n,t} - D_{n,t-1}\right) - \\ S_{n,leave} D_{n,t-1}\left(D_{n,t-1} - D_{n,t}\right) + \eta^{ch} P_{n,t}^{ch} \Delta t - \\ \dfrac{\eta^{ref} P_{n,t}^{dis} \Delta t}{\eta^{dis}}, \forall n \in N_j^{EV}, \forall t = T \end{cases} \quad (16)$$

Using Minkowski summation theory to summation, the dispatchable capacity calculation model of charging station is obtained, as shown in Eq. (17).

$$\begin{cases} P_{j,t}^{ch} = \sum_{n \in N_j^{EV}} P_{n,t}^{ch} \\ P_{j,t}^{dis} = \sum_{n \in N_j^{EV}} P_{n,t}^{dis} \\ S_{j,t} = \sum_{n \in N_j^{EV}} S_{n,t} \end{cases} \quad (17)$$

$$\begin{cases} P_{j,t}^{ch,\max} = \sum_{n \in N_j^{EV}} P_{\max,n}^{ch} D_{n,t}, \; P_{j,t}^{dis,\max} = \sum_{n \in N_j^{EV}} P_{\max,n}^{dis} D_{n,t} \\ S_{j,t}^{\min} = \sum_{n \in N_j^{EV}} S_n^{\min} D_{n,t}, \; S_{j,t}^{\max} = \sum_{n \in N_j^{EV}} S_n^{\max} D_{n,t} \\ \Delta S_{j,t} = \sum_{n \in N_j^{EV}} \left(S_{n,arrive} D_{n,t}\left(D_{n,t} - D_{n,t-1}\right)\right) \\ - \sum_{n \in N_j^{EV}} \left(S_{n,leave} D_{n,t-1}\left(D_{n,t-1} - D_{n,t}\right)\right) \end{cases} \quad (18)$$

$$\begin{cases} 0 \le P_{j,t}^{ch} \le P_{j,t}^{ch,\max}, \forall t \in T \\ 0 \le P_{j,t}^{dis} \le P_{j,t}^{dis,\max}, \forall t \in T \\ S_{j,t} = S_{j,t-1} + \Delta S_{j,t} + \eta^{ch} P_{j,t}^{ch} \Delta t - \dfrac{\eta^{ref} P_{j,t}^{dis} \Delta t}{\eta^{dis}}, \forall t \in T \\ S_{j,t}^{\min} \le S_{j,t} \le S_{j,t}^{\max}, \forall t \in T \end{cases} \quad (19)$$

Through the above process, the original large number and scattered EVs are aggregated into a whole and participate in the micro-energy grid modulation in the form of energy storage units.

(3) Prediction of the dispatchable potential of charging stations

In this paper, the schedulable potential $\beta_{\text{j}}$ is defined as:

$$\beta_j = \left\{P_{j,t}^{ch,\max}, P_{j,t}^{dis,\max}, \Delta S_{j,t}, S_{j,t}^{\min}, S_{j,t}^{\max}\right\} \quad (20)$$

### B. *The micro-energy network has recently optimized the scheduling*

(1) The micro-energy network has recently optimized the scheduling

In this paper, the actual value of new energy output is regarded as the sum of the predicted value and the random error, and the error follows the standard normal distribution of the variance of $\sigma_{\text{NEW}}$, that is:

$$P_{NEW}(t) \sim P_{NEW}^{f}(t) + N(0, \sigma_{NEW}) \quad (21)$$

(2) The objective function is optimized before the day

In the day-ahead stage, on the basis of considering the consumption of new energy and the emission of pollutants, the day-ahead optimal scheduling model is established with the optimization goal of minimizing the operating cost of the micro-energy grid.

$$
\begin{cases}
C_M = \min(C_G + C_{pollu} + C_{Gird} + \\
C_{ESS} + C_{EVC} + C_{DR} + C_{cur}) \\
C_G = \sum_{t=1}^{T}\left(\gamma_{GT}\Delta t + C_{GT}^{up}S_{GT}^{up}(t) + C_{GT}^{down}S_{GT}^{down}(t)\right) \\
\gamma_{GT} = aP_{GT}^3 + bP_{GT}^2 + cP_{GT} + d \\
C_{pollu} = \sum_{t=1}^{T}\left(P_{GT}(t)\cdot\Delta t\cdot K_{GT}^{pu}\right) \\
C_{Gird} = \sum_{t=1}^{T}\left(P_{Gird}(t)\cdot\Delta t\cdot\sigma_{Gird}\right) \\
C_{ESS} = \sum_{t=1}^{T}\left|P_{ESS}(t)\right|K_{ESS}\Delta t \\
C_{EVC} = \sum_{i=1}^{N_{EVC}}\sum_{t=1}^{T}c_{EVC}\left(P_{i,t}^{ch} + P_{i,t}^{dis}\right)\Delta t \\
C_{DR} = \lambda_e\sum_{t=1}^{T}\Delta L_{e,cut_t} + \lambda_h\sum_{t=1}^{T}\Delta L_{h,cut_t} \\
C_{cur} = \sum_{t=1}^{T}\left(P_{New_cur}(t)\cdot\Delta t\cdot\lambda_{cur}\right)
\end{cases} \tag{22}
$$

where $\gamma_{GT}$、 $C_{GT}^{up}$、 $C_{GT}^{\text{down}}$ denotes the fuel cost function of *GT*, start-stop cost; $S_{GT}^{up}(t)$、 $S_{GT}^{\text{down}}(t)$ indicating the start-stop state of GT; $K_{GT}^{\text{pu}}$、 $\sigma_{\text{Gird}}$、 $K_{ESS}$ , the cost factor of pollution per unit of power generation of *GT*, the interaction between the microgrid and the grid, and the loss of the energy storage system; $\lambda_e$ and $\lambda_h$ represent the unit compensation coefficient of electric and heat loa.

(3) Optimize scheduling constraints a few days ago

a Electrical power balance constraints

$$
\begin{cases}
P_{Load}(t) + P_{ES_ch}(t) + P_{Sell}(t) + P_{HP}(t) \\
= P_{NEW}(t) + P_{ES_dis}(t) + P_{GT}(t) + P_{buy}(t) \\
P_{ES_ch}(t) = P_{ESS_ch}(t) + \sum_{i=1}^{N_{EVC}} P_{i,t}^{ch} \\
P_{ES_dis}(t) = P_{ESS_dis}(t) + \sum_{i=1}^{N_{EVC}} P_{i,t}^{dis}
\end{cases} \tag{23}
$$

The opportunity constraint is used to convert Eq. (23) into the electric power balance constraint under the uncertainty of new energy:

$$
\begin{aligned}
&\eta \le P_r(P_{Load}(t) + P_{ES_ch}(t) + P_{Sell}(t) + P_{HP}(t) - \\
&P_{NEW}(t) - P_{ES_dis}(t) - P_{GT}(t) - P_{buy}(t) \ge 0)
\end{aligned} \tag{24}
$$

Where $\theta^{-1}(\eta)$is the inverse function of the standard normal distribution.

b Thermal power balance constraints

$$
\begin{cases}
Q_{HP_}t + H_{HS_dis_t} = L_{h_t} + H_{HS_ch_t} \\
Q_{Hp_t} \le Q_{Hp_\max} \\
H_{HS_ch_\min} \le H_{HS_ch_t} \le H_{HS_ch_\max} \\
H_{HS_dis_\min} \le H_{HS_dis_t} \le H_{HS_dis_\max}
\end{cases} \tag{25}
$$

In addition to the constraints mentioned above, the micro-energy network optimization model also needs to meet the constraints of each component in the previous subsection and the constraints on the charging station.

*C. Intra-day rolling optimization scheduling of micro-energy network*

(1) Scroll the time-domain method

The rolling time-domain method consists of the following basic steps:

a) Initialization: Set the initial state, the length of the time window, and optimize the objective function;

b) Optimization: In the current time window, based on the current system state, the optimization problem is solved to obtain the optimal control sequence;

c) Implement control: only implement the first control input in the optimal control sequence into the system;

d) Update: After the system status is updated, the time window rolls forward and the optimization and control implementation steps are repeated.

(2) Intra-day optimization model of micro-energy network

In the intra-day rolling optimization model, the goal is to minimize the amount of adjustment relative to the day-ahead scheduling instruction, and the minimum adjustment cost is taken as the objective function, and the optimization scheduling scheme is continuously adjusted according to the intra-day prediction data.

$$
C_{total} = \min(C_g + C_h) \tag{26}
$$

where $C_{total}$、 $C_g$、 $C_h$ represent the total adjustment cost, the adjustment cost of each part of the power system, and the adjustment cost of each part of the thermal system, respectively.

$$
C_g = \min\begin{bmatrix} \Delta P_{ESS}\sigma_{ESS} + \Delta P_{GT}\sigma_{GT} + \Delta P_{Gird}\sigma_{Gird} \\ +\Delta P_{EVC}c_{EVC} \end{bmatrix} \tag{27}
$$

$$
\begin{cases}
\Delta P_{ESS} = \sum_{t=1}^{T}\left\|P_{ESS}(t)\right| - \left|P_{ESS_0}(t)\right\| \\
\Delta P_{GT} = \sum_{t=1}^{T}\left\|P_{GT}(t)\right| - \left|P_{GT_0}(t)\right\| \\
\Delta P_{ESS} = \sum_{t=1}^{T}\left\|P_{Gird}(t)\right| - \left|P_{Gird_0}(t)\right\| \\
\Delta P_{EVC} = \sum_{i=1}^{N_{EVC}}\sum_{t=1}^{T}\left\|P_{i,t}^{ch} - P_{i,0,t}^{ch}\right| - \left|P_{i,t}^{dis} - P_{i,0,t}^{dis}\right\|
\end{cases} \tag{28}
$$

where $\sigma_{ESS}$、 $\sigma_{GT}$ are the adjustment coefficients of the battery, *GT*, and new energy.

$$
C_h = \min\left[\Delta P_{HS}\sigma_{HS} + \Delta P_{HP}\sigma_{HP}\right] \tag{29}
$$

$$
\begin{cases}
\Delta P_{HS} = \sum_{t=1}^{T}\left\|P_{HS}(t)\right| - \left|P_{HS_0}(t)\right\| \\
\Delta P_{HP} = \sum_{t=1}^{T}\left\|P_{HP}(t)\right| - \left|P_{HP_0}(t)\right\|
\end{cases} \tag{30}
$$

(3) Intra-day optimization scheduling constraints

In addition to the constraints mentioned in the day-to-day scheduling, the micro gas turbine ramp-up constraints need to be considered in the process of optimizing the scheduling of the ultra-short time scale within the day, and the constraints are as follows:

$$-R_{GT_d} \le P_{GT_0}(t) - P_{GT_0}(t-1) \le R_{GT_u} \qquad (31)$$

## IV. CASE ANALYSIS USING

In this chapter, the micro-energy grid system in Chapter Ⅱ is selected for simulation verification. The forecast data of wind power, photovoltaic, electric load and heat load in the day and day are shown in the figure below.

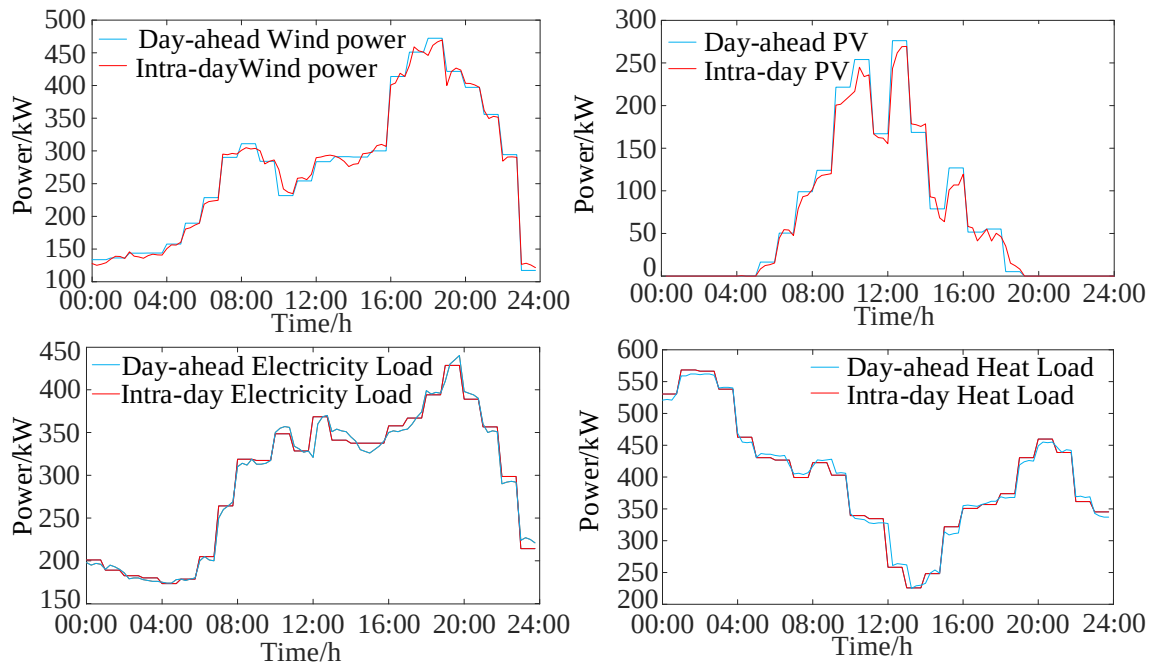

Fig. 2. Day-ahead and intra-day Forecasting Data

### A. *Analysis of day-ahead scheduling results*

Firstly, this paper obtains the prediction data of the dispatchable potential of each electric vehicle charging station through historical data, taking charging station 2 as an example, its dispatchable potential is shown in the figure below.

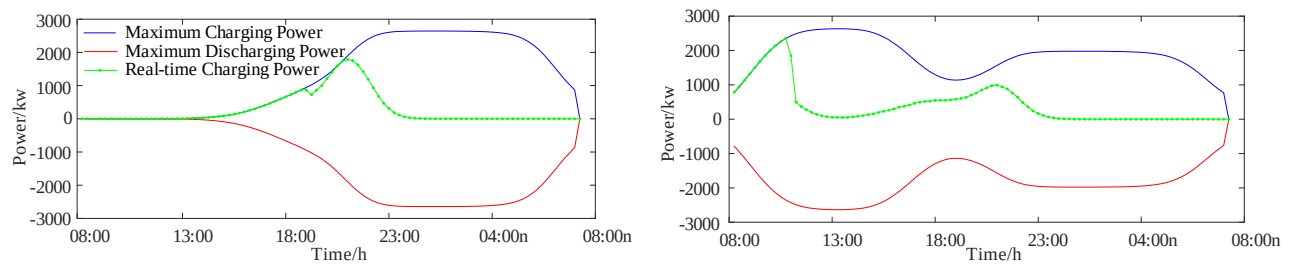

Fig. 3. Day-ahead Dispatchable Potential Forecast for Charging Station 2

The EV dispatchable potential prediction diagram is a set of envelopes that show that the intra-day charging power and amount of power of the charging station are in its controllable area, indicating that the calculation method in this paper is feasible. At the same time, as can be seen from Fig. 4, the dispatchable potential of the four charging stations is different at different times, and they cooperate with each other to provide a strong boost for the optimal scheduling of micro-energy networks as a generalized energy storage system.

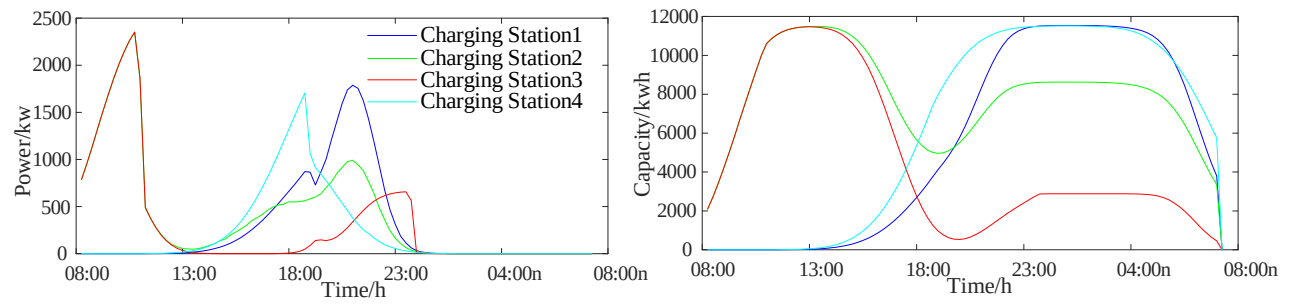

(a) Charging Power of Each Charging Station

(b) Quantity Changes in Charging Stations

Fig. 4. Comparison of Dispatchable Potential Among Charging Stations 2

Establish the following two scenarios for comparative analysis:

Scenario 1: Ignoring demand response

Scenario 2: Considering demand response

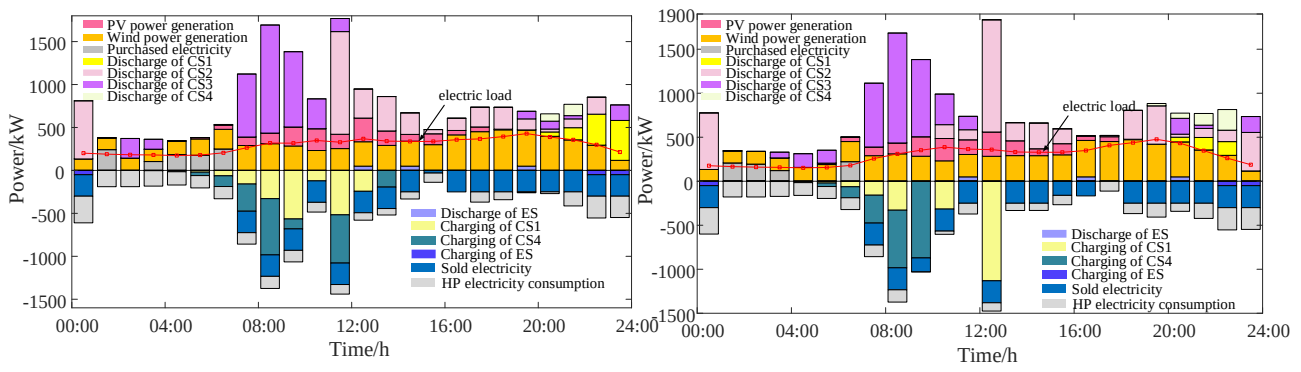

Fig. 5. Power Balance Results of Electrical Load for Scenario 1、2

During peak hours (10:00-12:00, 18:00-21:00), through electricity price guidance and compensation incentives, part of the load is actively interrupted, and the load can be transferred to the trough period of 1:00-9:00, 13:00-17:00 and 22:00-24:00. As a flexible energy storage unit, the charging station participates in peak shaving and valley filling in different time periods:10:00-12:00: Charging stations 2 and 3 are discharged in energy storage mode to relieve the load pressure on the power grid; 18:00-21:00: Charging station 1-3 co-discharge peak shaving; Low load period: The charging station switches to charging mode to absorb the surplus power of the power grid.

After cooperative scheduling, the peak-to-valley difference of the power grid is reduced by 18.5%, which effectively smooths out the load fluctuation and enhances the robustness of the system to the prediction uncertainty.

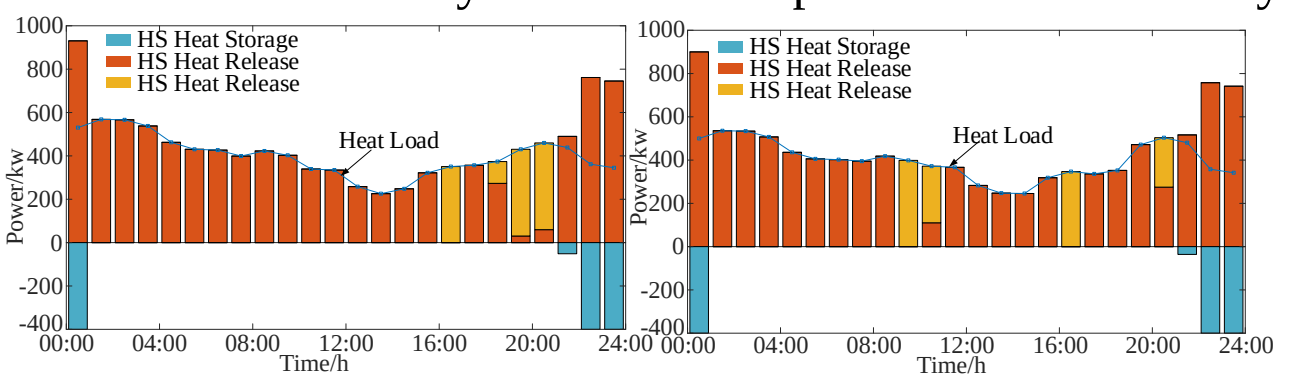

Fig. 6. Power Balance Results of Heat Load for Scenario 1、2

Since the heat source in the system is only the heat pump, there is a coupling relationship between electric heat, and the heat supply of the heat pump changes significantly after considering the dispatchable potential of electric vehicles and the demand response.

### B. *Intra-day scheduling results are analyzed*

The figure below shows only the day-ahead and intra-day forecasts for the maximum dispatchable potential of each charging station. It can be seen that the day-ahead and intra-day dispatchable potential show similar morphological characteristics.

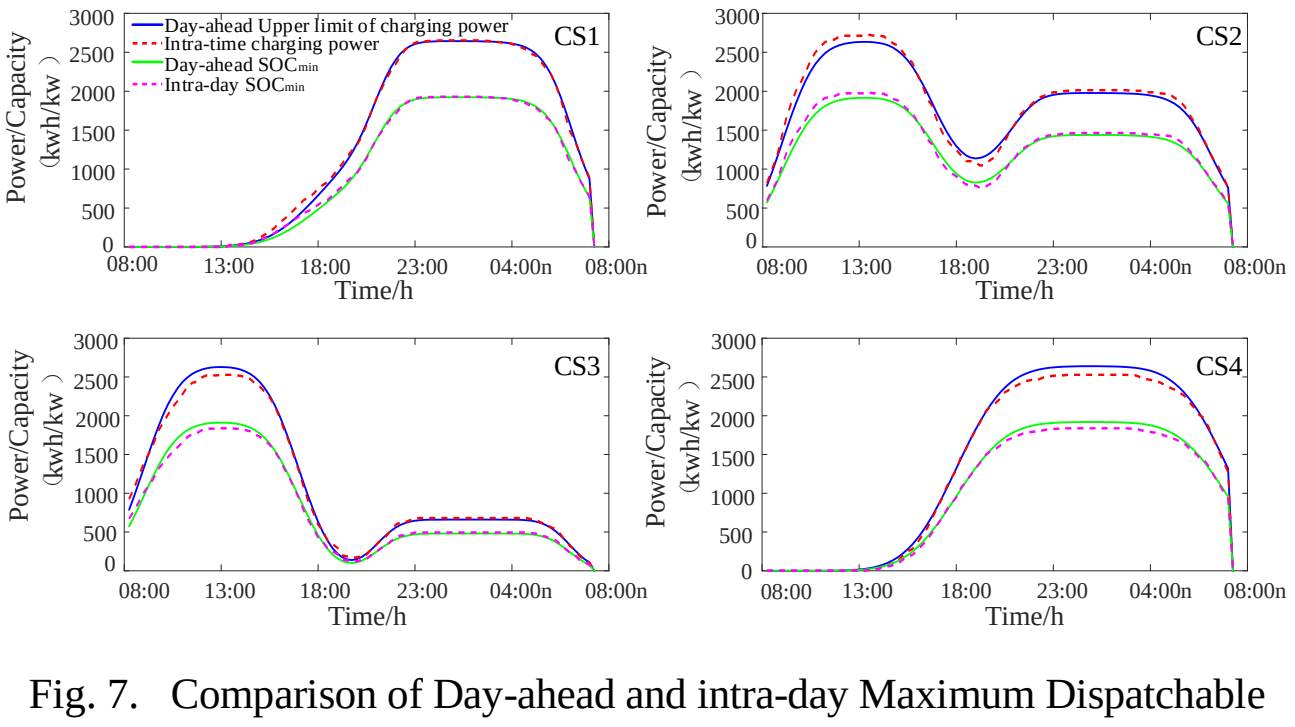

Fig. 7. Comparison of Day-ahead and intra-day Maximum Dispatchable Potential Among Charging Stations

As can be seen from Fig. 7, the actual output power of the dispatchable equipment in the day is basically consistent with the change trend of its day-ahead output plan, and in the day, various dispatchable equipment is adjusted on the basis of the day-ahead output plan to ensure the stability of the system and improve the robustness of the system in the unpredictable environment.

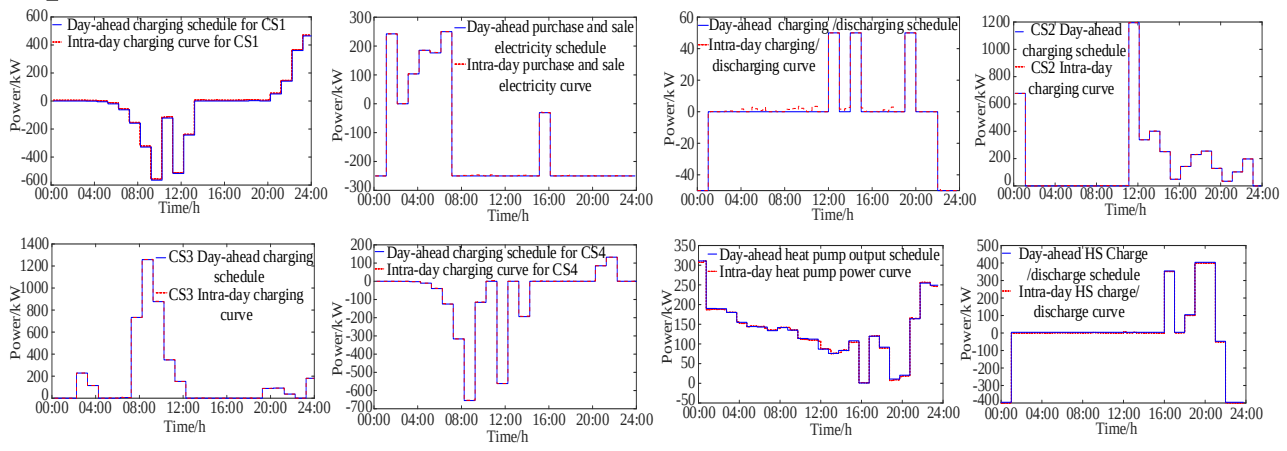

Fig. 8. Comparison of Day-ahead Generation Plan and Intra-day Output for Dispatchable Devices

The figure below shows the difference between the predicted value of wind and solar output and the actual value of wind and solar output of the two strategies proposed in this paper. There is a shortage of output in the day-to-day scheduling plan, while there is only wind and solar curtailment in intra-day scheduling. In order to avoid the shortage of output, a large amount of reserve capacity needs to be reserved, which indirectly leads to an increase in curtailment of wind and solar power. The two-stage optimal scheduling can track the output of new energy sources in real time, dynamically reduce the demand for spare capacity, and improve the space for new energy consumption.

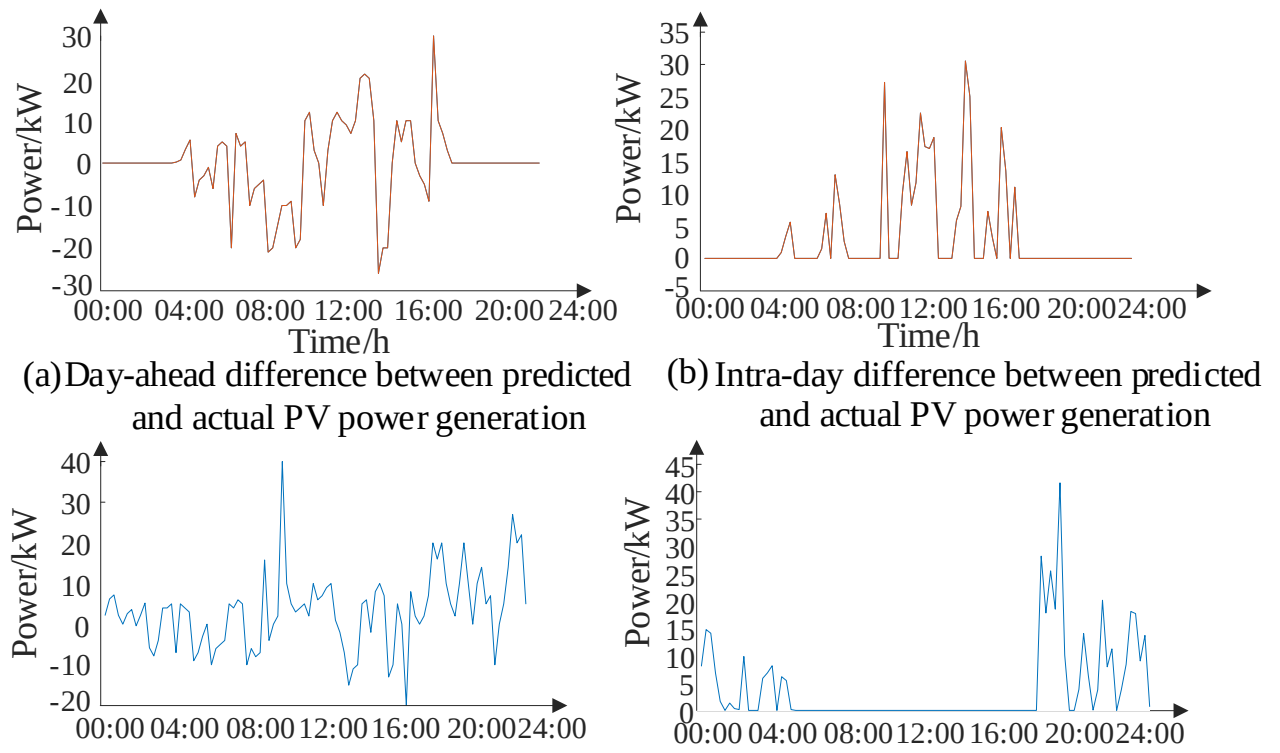

(a) Day-ahead difference between predicted and actual PV power generation

(b) Intra-day difference between predicted and actual PV power generation

(c) Day-ahead difference between predicted and actual wind power generation

(d) Intra-day difference between predicted and actual wind power generation

Fig. 9. Day-ahead and intra-day difference between predicted and actual renewable energy output

In the electricity market, the actual output is lower than the planned value (negative deviation) usually faces the deviation assessment cost, the following table is a comparison of the deviation cost under the two strategies, it can be seen that the day-ahead dispatch plan due to the existence of output shortage, need to pay the deviation assessment fee, the cost is higher. By avoiding the shortfall in output, intra-day adjustment significantly reduces economic penalties and operating costs, and improves the system's economics in unpredictable forecasts

TABLE I. COMPARISON RESULTS OF EMERGENCY ELECTRICITY PURCHASE COSTS UNDER TWO STRATEGIES

| Assessment index | Strategy 1 | Strategy 2 |
|---|---|---|
| **Cost variance of wind power/YUAN** | 224.4 | 0 |
| **Cost variance of solar power/YUAN** | 311.3 | 0 |

## V. SUMMARY

In this paper, we propose a rolling optimization scheduling method considering the multi-time scale rolling optimization of electric vehicles connected to micro-energy networks. The conclusions are as follows:

(1) Based on the Minkowski and the theory, the dispatchable capacity of different charging stations is effectively calculated, and its potential is fully utilized, and the proposed method effectively reduces the daily operating cost of the distribution network, promotes the consumption of new energy, and improves the operation economy of the system through simulation verification.

(2) The demand response mechanism of the flexible load on the user side is introduced, and the incentive compensation measures are used to give full play to the flexible adjustment ability of the load side, and the simulation verifies that the method effectively suppresses the fluctuation of new energy output and realizes peak shaving and valley filling;

(3) The two-stage optimization scheduling strategy of the micro-energy network proposed in this paper combines the day-ahead global optimization with the intra-day rolling correction, which effectively improves the economy and robustness of the system in the unpredictable prediction environment. At the same time, this method significantly reduces the economic penalty and operating cost by avoiding the lack of output, and improves the economy of the system in the unpredictable environment.

## ACKNOWLEDGMENT

This work is supported by the State Grid Liaoning Electric Power Co., LTD. Science and Technology Project(No.2024YF-74), Research on key technologies of rural micro-energy network optimization control to support electric vehicles going to the countryside.